# Unsaturated magnetoconductance of epitaxial La$_{0.7}$Sr$_{0.3}$MnO$_3$ thin films in pulsed magnetic fields up to 60 T

Wei Niu,[1] Xuefeng Wang,[1,a] Ming Gao,[1] Zhengcai Xia,[2] Jun Du,[3] Yuefeng Nie,[4] Fengqi Song,[3] Yongbing Xu[1] and Rong Zhang[1]

[1] *National Laboratory of Solid State Microstructures, Collaborative Innovation Center of Advanced Microstructures, and School of Electronic Science and Engineering, Nanjing University, Nanjing 210093, China*

[2] *Wuhan National High Magnetic Field Center, Huazhong University of Science and Technology, Wuhan 430074, China*

[3] *National Laboratory of Solid State Microstructures, Collaborative Innovation Center of Advanced Microstructures, and School of Physics, Nanjing University, Nanjing 210093, China*

[4] *College of Engineering and Applied Sciences, Nanjing University, Nanjing 210093, China*

We report on the temperature and field dependence of resistance of La$_{0.7}$Sr$_{0.3}$MnO$_3$ thin films over a wide temperature range and in pulsed magnetic fields up to 60 T. The epitaxial La$_{0.7}$Sr$_{0.3}$MnO$_3$ thin films were deposited by laser molecular beam epitaxy. High magnetic field magnetoresistance curves were fitted by the Brillouin function, which indicated the existence of magnetically polarized regions and the underlying hopping mechanism. The unsaturated magnetoconductance was the most striking finding observed in pulsed magnetic fields up to 60 T. These observations can deepen the fundamental understanding of the colossal magnetoresistance in manganites with strong correlation of transport properties and magnetic ordering.

Since the observation of colossal magnetoresistance (CMR) in doped manganite perovskites,[1] there has been a recent surge of research in order to understand the strong correlation of resistance and magnetic ordering in these materials.[2-4] The strong coupling between electrical and magnetic orderings leads to the sophisticated physical phases in manganites.[3-5] However, the magnetotransport phenomena in manganites have been mainly studied in relatively weak magnetic fields.[2,6,7] Only few papers reported the high-field resistivity of La/Sr based manganites.[8-10] In order to deepen knowledge of the specific physical features of mixed valence manganites, applying the high pulsed magnetic fields up to 60 T is a powerful method to elucidate the magnetotransport processes in extreme conditions.[8]

---

[a] Author to whom correspondence should be addressed. Electronic mail: xfwang@nju.edu.cn.

Appling the magnetic fields enhances the ferromagnetic phase and then reduces the spin scattering and produces the CMR effect in doped manganites.[10] In manganites, the conductivity is supposed to emerge from hopping of magnetic polarons above Curie temperature ($T_C$), and from the double exchange mechanism below $T_C$.[11] The analytical dependence of the CMR effect versus magnetic fields has not yet been fully understood, but correlations between magnetization and resistivity suggest an important role of the polaron hopping below $T_C$.[12] Concerning the important issue of the coupling between the electrical and magnetic orderings, various mechanisms and models have been proposed.[10,13] Moreover, the debates on the transport properties of manganites still remain controversial.

In this work, we deposit $La_{0.7}Sr_{0.3}MnO_3$ (LSMO) films by laser molecular beam epitaxy (LMBE) on $SrTiO_3$ (STO) substrates. In order to get the deeper insight, we present an experimental approach for magnetoresistance measurements in pulsed magnetic fields up to 60 T and discuss the transport and magnetic properties of epitaxial LSMO films measured between 2 and 300 K. It is found that the field dependence of magnetoresistance can be fitted by the Brillouin function in the ferromagnetic region without the matter of orientation of magnetic fields, suggesting the existence of magnetically polarized regions and the underlying hopping mechanism. Moreover, the most striking finding is the magnetoconductance does not saturate up to 60 T.

In order to avoid extrinsic magnetoresistance from grain boundaries, the experiments discussed in this work are performed on epitaxial films, which offer the advantage of nearly perfect quality of single crystal, thus close to the intrinsic properties of the LSMO. Furthermore, thin films represent the most feasible sample form for the potential device applications and the subsequent fabrication of multilayers. In brief, LSMO films were grown on $TiO_2$-terminated STO (001) substrates using LMBE technique by applying a KrF excimer laser at a repetition of 2 Hz. The temperature and oxygen pressure during the film growth were maintained at 750 ºC and $2\times10^{-3}$ mbar with 20% ozone, respectively. With the help of ozone, the epitaxial thin films have the better quality with no need of post-annealing. The detailed fabrication process and characterization of the epitaxial LSMO films were described elsewhere.[14] The crystalline structure of the samples were examined by X-ray diffraction (XRD) using Bruker D8-Discover. The surface morphology of the film was investigated by a high-resolution atomic force microscopy (AFM, Asylum Cypher under ambient conditions). Magnetic properties were measured with a superconducting quantum interference device (SQUID-VSM). The resistivity was measured using a standard four-probe method. The low-field transport properties were measured by a physical property measurement system (PPMS-9T, Quantum Design) and high-field measurements were performed in a pulsed magnetic fields (up to 60 T) at Wuhan National High Magnetic Field Center, China. Particular attention should be paid to verify the absence of Joule and eddy current heating effects, which otherwise would lower the signal to noise ratio or even conceal the true signal.[15]

Figure 1(a) shows the *in-situ* reflection high-energy electron diffraction (RHEED) intensity oscillations of the grown film, indicating that the growth proceeds in an ideal layer-by-layer mode. The peaks of the RHEED oscillations represent the growth of exact unit cell (u.c.)-control thickness. The left inset of Fig. 1(a) is the typical RHEED pattern of STO (001) substrate prior to deposition at 750 °C, while the sharp streaky line in the right inset is characteristic of LSMO RHEED pattern after deposition. An atomically smooth surface with clear steps is evidenced in a typical AFM image [upper of Fig. 1(b)] and the corresponding AFM line profile [bottom of Fig. 1(b)] across the terraces shows that the average terrace height is about 0.4 nm. The step height corresponds to the lattice constant of LSMO, indicating a well-defined, atomically flat surface. Fig. 1(c) presents the XRD pattern of the LSMO film. The epitaxial thin film has no extra phases and is highly *c*-axis oriented. Only (00l) LSMO || (00l) STO in plane orientation is observed. Although the film is ultrathin (20 u.c., less than 8 nm), its diffraction signals can still be clearly seen, which further proves the nearly perfect single-crystal structure of our high-quality LSMO films.

In order to obtain a meaningful comparison between electrical and magnetic properties, the transport measurements are performed on a strip cut from the same film that is also used for the magnetic measurements. Figure 2(a) shows the resistivity versus temperature curve in the range of 2–400 K for the sample. As the temperature decreases from 400 K in the paramagnetic phase, the resistivity initially increases, reaching the maximum at the metal-insulator transition temperature ($T_p$) ~340 K. The $T_p$ value is consistent with its Curie temperature ($T_C$) which is thickness-dependent as determined by the temperature-dependent magnetization curve [the black square in Fig. 2(a)]. Here $T_p$ is close to the $T_C$ of the bulk LSMO ($T_C$ ~ 369 K),[16] further showing the high quality of our LSMO films. In the ferromagnetic phase ($T < T_C$), the LSMO ultrathin film exhibits a typical metallic behavior down to low temperatures. Interestingly, the resistivity does not show a residual resistance behavior when $T \to 0$, but reaches a minimum at temperature around 25 K and increases until the temperature decreases to 2 K. This is attributed to the quantum interference effect in epitaxial thin films.[14] Figure 2(b) shows the clear ferromagnetic hysteresis loops of LSMO thin film at different temperatures with the in-plane magnetic field from -5000 Oe to 5000 Oe. The saturation magnetization of LSMO thin film without post-annealing is analogous to the result of Yuan *et al*. reported previously.[17]

The high magnetic field increases sinusoidally to the maximum field within 10 ms. Systematic errors are overcome by normalization to the zero-field value $\rho_0$. For clarity, only a few representative plots are given here. Since the sample temperature is restricted to $T < 300$ K for the pulsed field cryostat, we focus on the magnetoresistance in the ferromagnetic region. Figure 3(a) shows the field-dependent magnetoresistance of LSMO film with the magnetic field parallel to the current applied on the sample. A striking correspondence is found between the negative magnetoresistivity and the Brillouin function $B_J$ for temperature below $T_C$. Brillouin function $B_J$ is expressed as

$$\Delta\rho/\rho_0 = A(T)B_J\left[g\mu_B J(T)H/k_B T\right] \qquad (1)$$

where $A(T)$ is a coefficient, $g = 2$ is gyromagnetic ratio, $\mu_B$ is the Bohr magneton, $J(T)$ is the average spin moment at the hopping sites, $H$ is the external magnetic field, $k_B$ is the Boltzmann constant, and $T$ is the temperature. Fitting the data according to $\Delta\rho/\rho_0=A(T)B_J$ in the ferromagnetic state gives the fitting parameter $J(T)$. As can be seen from Fig. 3 (a), the fits are of good quality described in lines. In the ferromagnetic phase, the field induces resistivity to decrease, which scales with the Brillouin function, suggesting the existence of magnetically polarized regions and the transport process of hopping. Most of magnetoresistance data in previous reports of epitaxial films in pulsed magnetic fields were also acquired from the case of current parallel to magnetic fields.[11,15,18,19] Additionally, we measure the field dependence of magnetoresistance with the magnetic field perpendicular to the surface of the film. The data are plotted in Fig. 3(b). The nice fits can be also applied by the Brillouin function, indicating that the hopping mechanism is independent of the magnetic-field orientation.

The temperature dependence of the spin moment $J$ for the LSMO film is shown in Fig. 3(c) (plotted with a log-log scale). At low temperatures, $J$ is smaller than the theoretical value of 4 $\mu_B$ for the bare $Mn^{3+}$, which might be related to the metallic nature of charge transport.[18] $J$ increases logarithmically with increasing temperature below $T_C$. The large value of $J$ suggests the formation of magnetic clusters, which is in accordance with the previous report.[15] The $J$-$T$ curve suggests that the size of the polaron increases with increasing temperature. The average spin moments, $J_\parallel$ and $J_\perp$, under different magnetic orientation display the similar behavior, further indicating that the hopping mechanism of the epitaxial thin films is independent of the magnetic-field orientation.

Applying magnetic fields, the resistance of common conductor generally exhibits a quadratic dependence at the low fields which saturates at high magnetic fields.[20] However, many examples show unsaturated behaviors in various materials.[20-22] Previous reports show that polycrystals with grain boundaries exhibit a linear and unsaturated field dependence of the high-field magnetoconductance in the whole ferromagnetic region.[23-25] Figure 3(d) shows the magnetoconductance ($G$) of the epitaxial LSMO film, $G$ is normalized to the value obtained in zero-field $G_0$ for eliminating the systematic error. The linear magnetoconductance behaviors in polycrystals are normally attributed to the strong antiferromagnetic coupling of the spins at grain boundaries.[9,24] Nevertheless, the epitaxial single-crystal film shows a different tendency: The conductance increases with the magnetic field in a slightly linear manner, then it deviates the linear behavior above 20 T. It still keeps unsaturated in pulsed magnetic fields up to 60 T. This interesting unsaturated magnetoconductance is originated from the CMR nature of the manganites.

In conclusion, single-crystal LSMO films with atomically flat surface have been epitaxially grown by LMBE. We present an experimental approach for magnetoresistance measurements in pulsed magnetic fields and discuss the resistance data measured in magnetic fields up to 60 T. The field dependence of the magnetoresistance can be fitted by the Brillouin function in the ferromagnetic phase without the matter of orientation of magnetic fields, suggesting the existence of magnetically polarized regions and the hopping mechanism. A hopping probability of conduction electrons is known to be strongly related to mutual orientation of magnetic moments of neighboring manganese ions. Thus, the transport processes of conduction electrons become spin dependent, which may find the potential applications in spin valves and magnetic tunnel junctions. Moreover, the unsaturated magnetoconductance is observed in pulsed magnetic fields up to 60 T. These observations can deepen the understanding of the CMR in manganites with strong correlation of transport properties and magnetic ordering.

This work was supported by the National Key Projects for Basic Research of China (Grant No. 2014CB921103), the National Natural Science Foundation of China (Grant No. 11274003), and the PAPD project.

**Figures**

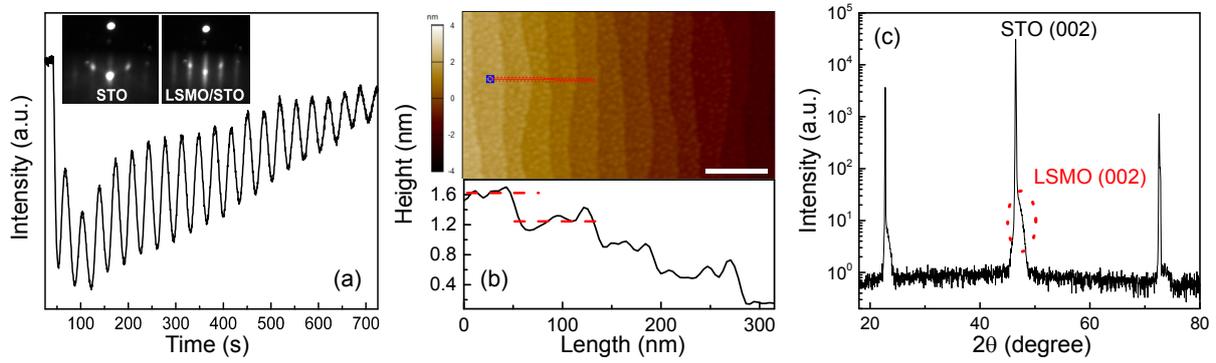

**FIG. 1. Structural analysis of the epitaxial LSMO films grown on STO.** (a) Typical RHEED intensity oscillation of 20 u.c. LSMO thin film. The inset shows the RHEED patterns of STO substrate before the growth (left) and LSMO film after the growth on STO (right). (b) A typical AFM image and the corresponding AFM line profile across the terraces. The scale bar is 200 nm. (c) XRD $\theta$-$2\theta$ scan of the 20 u.c. LSMO film. The lump on the right of STO diffraction peak is the contribution from the LSMO thin film.

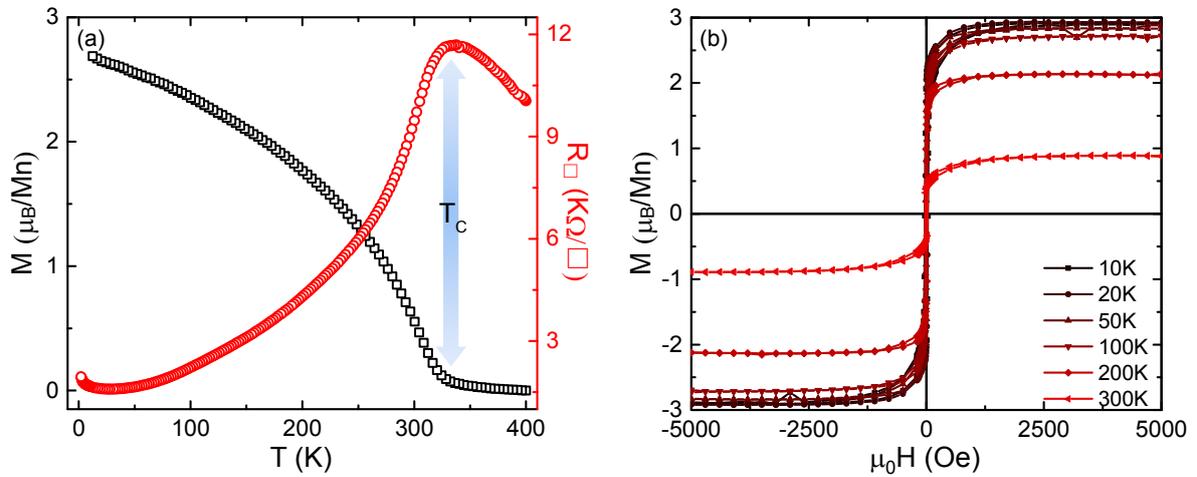

**FIG. 2. Magnetic properties of 20 u.c. LSMO film.** (a) Temperature-dependent resistivity and magnetization. The arrows indicate the Curie temperature, $T_C$. (b) Magnetic hysteresis loops measured in plane at different temperatures. The diamagnetic contribution of substrate has been subtracted.

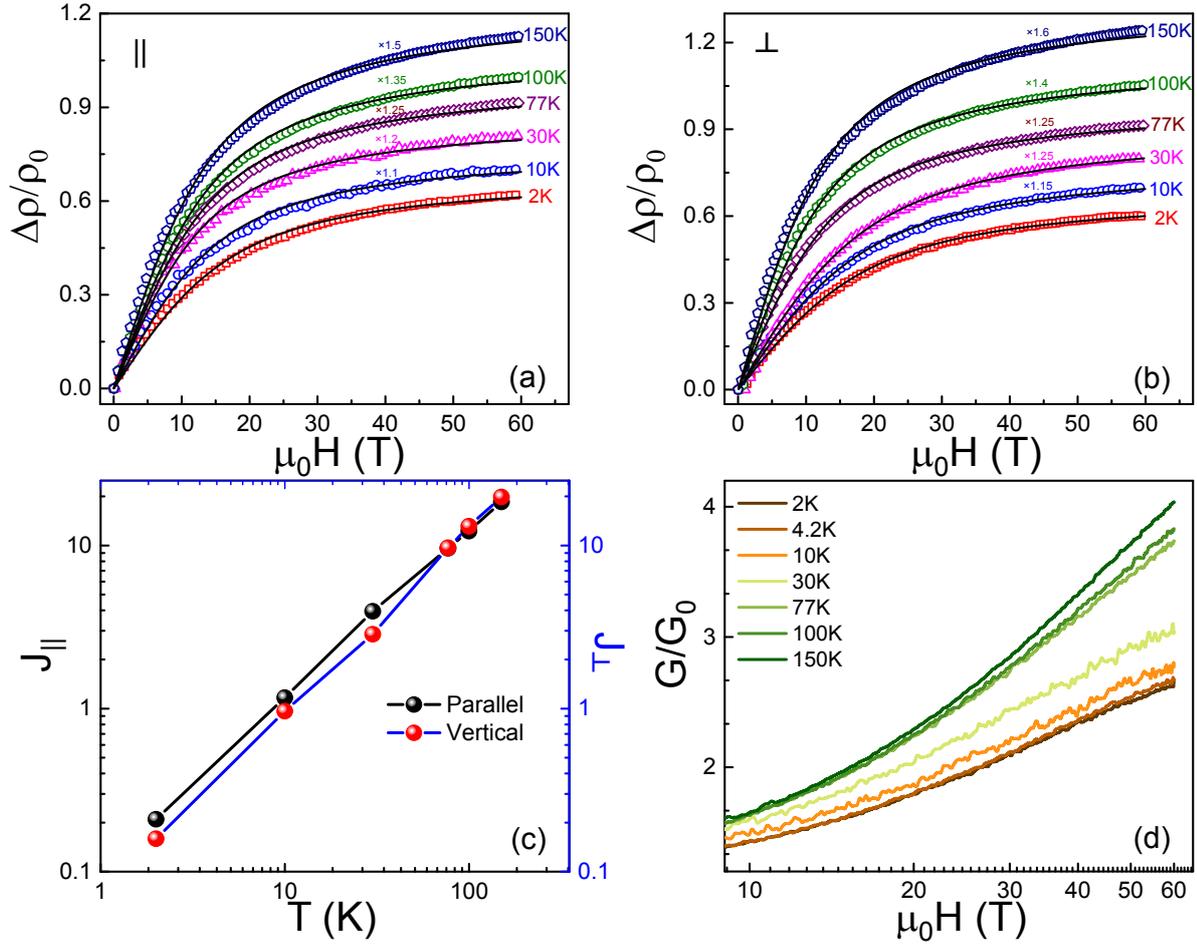

**FIG. 3. Transport properties of 20 u.c. LSMO film in pulsed magnetic fields up to 60 T.** (a),(b) Field-dependent magnetoresistance of epitaxial LSMO film measured under the magnetic fields which are in-plane ($\parallel$) and out-plane ($\perp$), respectively. (c) Temperature dependence of the effective spin moment $J$ for LSMO film. (d) Field dependence of $G$ measured for epitaxial LSMO film at various temperatures, and normalized to the zero field value $G_0$, showing unsaturated behavior up to 60 T.